\documentclass[12pt]{article}
\usepackage{graphicx}
\usepackage{amssymb}

\def\D{\Delta}
\def\h{\hbar}
\def\B{\beta}
\def\A{\alpha}

\title{\LARGE{Microscopic black hole and uncertainty principle with minimal length and momentum}}
\author{M. M. Stetsko\footnote{E-mail: mstetsko@gmail.com, mykola@ktf.franko.lviv.ua}\
\\
  {\small Department of Theoretical Physics, Ivan Franko National University of Lviv,}\\
{\small 12 Drahomanov Str., Lviv, UA-79005, Ukraine
         }}

\begin{document}
\maketitle

\abstract{We investigate a microscopic black hole in case of
modified generalized uncertainty principle with a minimal
uncertainty in position as well as in momentum. We calculate
thermodynamical functions of a Schwarzschild black hole such as
temperature, entropy and heat capacity. It is shown that
incorporation of minimal uncertainty in momentum leads to minimal
temperature of a black hole. Minimal temperature gives rise to
appearance of a phase transition. Emission rate equation and black
hole's evaporation time are also obtained.}



\section{Introduction}
Quantum black holes have been intensively investigated for recent
years. This interest was mainly stimulated by the development of
high energy physics. One of the most promising open issues of high
energy physics is the development of theories with extra
space-like dimensions. It was proposed that extra space-like
dimensions can lower the Planck scale to the TeV region
\cite{arkani-hamed, arkani-hamed_2, antoniadis, Randall99}. It is
expected that particle collisions with center-of-mass energy above
the Planck scale and impact parameter smaller than the horizon
radius can be a source for producing micro black holes and branes
\cite{BH_production, Eardley, Yoshino, Yoshino_2, Berti,
Cavaglia_Jhep, Kanti_09}. (Some critical remarks to this scenario
can be found in \cite{Voloshin, Rychkov04}). So the energy
sufficient for producing of micro black holes can in principle be
achievable even for present generation of particle accelerators.
The lowering of the Planck scale possibly allows to by-pass the
hierarchy problem. Identification of the ultraviolet cutoff scale
with electroweak energy one $\Lambda_{EW}$ can lead to radiative
stability without using additional assumptions such as
supersymmetry or technicolor. It is considered that the
unification of gravity with other gauge interactions at
electroweak scales can be obtained in theories with extra
dimensions. Another important feature of theories with extra
dimensions is violation of Newton's law of gravitation $\sim1/r^2$
and this deviation can be possibly probed in the nearest feature.

Evolution of micro black holes, formed in high energy collisions
can be divided into several stages, so called phases: loss of
gauge charges and multipole momenta (balding phase), loss of
angular momentum (spin-down phase) and the last one is so called
Schwarzschild phase when a black hole loses energy via Hawking
radiation. The final stage of Schwarzschild phase is not well
understood due to the lack of fully-fledged theory of quantum
gravity. Several attempts to understand this stage were made with
the help of Generalized Uncertainty Principle (GUP) \cite{adler,
cavaglia04, scardigli_2}. It was shown that GUP prevents complete
evaporation of black holes leaving some remnants. Noncommutative
geometry was also applied to quantum black holes. The key point in
this approach is the noncommutativity inspired modification of
black hole's metrics \cite{Nicolini08}.

Another line of research that has renewed interest to micro black
holes, especially with GUP, is cosmology. Recent investigations of
the CMB power spectrum have brought an idea about preinflationary
epoch \cite{Powell, Wang}. The preinflationary scenario proposed
in \cite{scardigli} was based on the micro black hole production
and a set of first principles such as generalized uncertainty
principle and holographic principle. These principles allow one to
obtain self-consistent description of the suppression of the CMB
quadrupole without additional assumptions. As it was pointed out
that GUP can lead to the matter domination in the preinflatory
epoch. In spite of some success of matter dominant preinflation
this scenario has some drawbacks and needs further investigation.

This paper is organized as follows. In the second section we
consider modified generalized uncertainty principle with minimal
position and momentum and obtain an expression for the black
hole's temperature. In the third section we calculate entropy and
heat capacity of black hole. In the fourth section emission rate
relation is considered and black hole's evaporation time is
calculated. The fifth section contains conclusions and some
mathematical details are considered in Appendix.

\section{Uncertainty principle with minimal length and momentum and black hole's temperature}
Recent studies in string theory brought an idea about
generalization of Heisenberg uncertainty principle \cite{gross,
witten}. Similar suggestion was proposed to reconcile principles
of Quantum Mechanics and General Relativity \cite{maggiore,
Scardigli_99}. Generalized Uncertainty Principle (GUP) takes form:
\begin{equation}\label{GUP}
\D p\D x\geqslant\frac{\h}{2}(1+\B\D p^2)
\end{equation}
As it is known such a generalization leads to appearance of
minimal uncertainty in position, so called minimal length. It was
shown that gravity-induced decoherence gives rise to similar
modification of uncertainty principle but with minimal uncertainty
for momentum \cite{Kay, Kay_CQG, Kay_07}. These two assumptions
about minimal uncertainties in position and momentum might be
merged in order to obtain modified generalized uncertainty with
both minimal length and momentum. We can write uncertainty
relation:
\begin{equation}\label{MGUP}
\D x\D p\geqslant\frac{\h}{2}\left(1+\A\D x^2+\B\D p^2\right)
\end{equation}
It is easy to show that $\D x\geqslant\D
x_{min}=\h\sqrt{\B}/(\sqrt{1-\h^2\A\B})$ and $\D p\geqslant\D
p_{min}=\h\sqrt{\A}/(\sqrt{1-\h^2\A\B})$. We note that in order to
get these minimal uncertainties parameters $\A$, $\B$ must be
positive and $\h^2\A\B<1$. We also remark that uncertainty
principle (\ref{MGUP}) was introduced after consideration of a
gedanken experiment for the simultaneous measurement of position
and momentum of a particle in de Sitter spacetime \cite{Bambi}.

 The uncertainty relation (\ref{MGUP}) can be obtained if we suppose that
operators of position and momentum obey deformed commutation
relation:
\begin{equation}\label{al_osc}
[x,p]=i\h (1+\A x^2+\B p^2),
\end{equation}
where parameters of deformation $\alpha$ and $\beta$ are assumed
to be positive ($\alpha,\beta>0$) and $\h^2\A\B<1$. We note that
commutation relation (\ref{al_osc}) and uncertainty principle
(\ref{MGUP}) firstly appeared in quantum group investigations
\cite{kempf1}.

We point out that some troubles appear when one tries to
generalize the algebra (\ref{al_osc}) for a case of higher
dimension. Probably the main difficulty is to satisfy closedness
condition. It was proposed to generalize commutation relation
(\ref{al_osc}) in the following way \cite{Banerjee, mignemi}:
\begin{eqnarray}\label{SDS_alg}
[x_i,p_j]=i\h\left(\delta_{ij}+\A x_ix_j+\B
p_jp_i+\sqrt{\A\B}(p_ix_j+x_jp_i)\right).
\end{eqnarray}
Whereas other two commutation relations need to be written in a
proper way to fulfil the Jacobi identity. It was also proposed the
representation of operators obeying (\ref{SDS_alg}). We also point
out that classical commutation relations that use classical
counterpart of (\ref{SDS_alg}) are well defined (see Appendix).

Let us come back to the uncertainty (\ref{MGUP}) and rewrite it in
the form:
\begin{equation}
\frac{1}{\h\B}\left(\D x-\sqrt{\D
x^2\left(1-\h^2\A\B\right)-\h^2\B}\right)\leqslant\D
p\leqslant\frac{1}{\h\B}\left(\D x+\sqrt{\D
x^2\left(1-\h^2\A\B\right)-\h^2\B}\right)
\end{equation}
It is easy to see that right hand side inequality does not recover
ordinary uncertainty principle in the limit $\B\rightarrow 0$ or
(and) $\A\rightarrow 0$. So we obtain
\begin{equation}\label{final_MGup}
\D p\geqslant\frac{1}{\h\B}\left(\D x-\sqrt{\D
x^2\left(1-\h^2\A\B\right)-\h^2\B}\right)
\end{equation}

 It was proposed that applying uncertainty principle to a
black hole one can get Hawking temperature up to a some
``calibration'' factor \cite{adler}. As it was pointed out by
Adler and collaborators GUP prevents a black hole from complete
evaporation and temperature catastrophe. That idea attracted a lot
of attention and was generalized for the case of higher dimension
\cite{cavaglia04, scardigli_2, Myung}.

Since evaporated particles appear just outside of the horizon
surface it means that uncertainty in position for them can not be
less than linear size of a black hole. If we assume that the
uncertainty in position for a particle is less than linear size of
the black hole it means that particle emerges inside the black
hole and can not penetrate through the horizon surface. On the
other hand if we suppose that the uncertainty in position is
greater than linear size of the black hole then the emitted
particle appears not on the outside of horizon surface but
somewhere beyond it and as a consequence the temperature of the
black hole will be different than it should be.  To obtain
temperature of the black hole we equate uncertainty in position
for emitted particle to the double of Schwarzschild radius.
\begin{equation}\label{uncert-rad}
\D x=2R_S.
\end{equation}

 Having used (\ref{final_MGup}) and supposing that $\D
E=c\D p$ we obtain the uncertainty in the energy of emitted
particle.
\begin{equation}\label{uncert_energy}
\D E=c\D p\geqslant\frac{c}{\h\B}\left(2R_S-\sqrt{
4R^2_S\left(1-\h^2\A\B\right)-\h^2\B}\right)
\end{equation}

Let us suppose that we deal with ordinary Hawking effect and
consider Hawking radiation just outside the event horizon of a
Schwarzschild black hole. We identify the uncertainty of energy
$\D E$ given by the relation (\ref{uncert_energy}) with the
thermal energy of an emitted particle. So we can use well-known
energy-temperature relation $\D E=3T$ (We have used thermal energy
for photons and here we set Boltzmann constant equal to unity
$k_B=1$). Taking into account all these remarks and substituting
the mass $M$ of a black hole instead of its gravitational radius
$R_S$ ($R_S=2GM/c^2$) we obtain relation for the temperature of
emitted particles
\begin{equation}\label{BH_temp}
T=\frac{c}{\pi\h\B}\left(\frac{4GM}{c^2}-\sqrt{\frac{16G^2M^2}{c^4}\left(1-\h^2\A\B\right)-\h^2\B}\right)
\end{equation}
We remark that in the latter relation we have used ``calibration''
factor $\pi$ instead of $3$ in order to get correct relation for
Hawking temperature when parameters of deformation tend to zero
($\A,\B\rightarrow 0$). Probably the correct result for the
Hawking temperature can be reproduced if one uses proper
equipartition law on the horizon surface or just outside it. It
was shown that number of degrees of freedom on the horizon surface
is defined by its surface area \cite{Padmanabhan}. So it is worth
expecting that proper equipartition law leads to correct
``calibration'' factor. It should be noted that Hawking
temperature for the Schwarzschild-AdS black hole with modified GUP
(\ref{MGUP}) was obtained in \cite{Park08}. But in order to
reproduce correct relation for the temperature of the black hole
one of the deformation parameters was held fixed \cite{Park08}.
Unfixing this parameter and taking Schwarzschild metrics instead
of Schwarzschild-AdS one we arrive at the relation
(\ref{BH_temp}). We point out that uncertainty relation
(\ref{MGUP}) can be used with Schwarzschild metrics because it can
be derived without any assumption about the de Sitter spacetime.
We also remark that temperature of Schwarzschild-AdS black hole
with generalized uncertainty principle that leads to appearance of
minimal momentum was firstly considered in \cite{BolenGERG}.

It was pointed out \cite{BolenGERG} that uncertainty principle
does not describe the origin of such effects as semiclassical wave
scattering or particle tunnelling but only their consequence on
the measurement process. To explain the origin of black hole's
evaporation it is necessary to know the quantum states of the
black hole from which the exact uncertainty principle follows. It
was suggested that black holes thermodynamics is a low-energy
effect of small-scale physics. It is known that any theory of
quantum gravity contains some kind of uncertainty principle that
reduces to Heisenberg principle at low energies. So black hole
thermodynamics should not depend too much on the details of
underlying quantum gravity theory \cite{BolenGERG}. This is in
agreement with Visser's conclusion that Hawking radiation requires
ordinary quantum mechanics and slowly evolving future horizon, so
to explain black hole thermodynamics quantum gravity principles
are not necessary \cite{visser}.

 Uncertainty relation (\ref{MGUP}) gives rise to the existence
of a minimal mass of a black hole:
\begin{equation}
M_{min}=\frac{\h c^2\sqrt{\B}}{4G\sqrt{1-\h^2\A\B}}=\frac{
cM^2_{Pl}\sqrt{\B}}{4\sqrt{1-\h^2\A\B}},
\end{equation}
where $M_{Pl}=\sqrt{\h c/G}$ is the Planck's mass. So according to
the relation (\ref{BH_temp}) it leads us to the finite temperature
at the final point of Hawking radiation, which takes following
form:
\begin{equation}
T_{final}=\frac{c}{\pi\sqrt{\B}\sqrt{1-\h^2\A\B}}
\end{equation}

One can see that uncertainty relation (\ref{MGUP}) causes
increasing of black hole's temperature in comparison with relation
(\ref{GUP}).

\begin{figure}
  \centerline{\includegraphics[scale=1,clip]{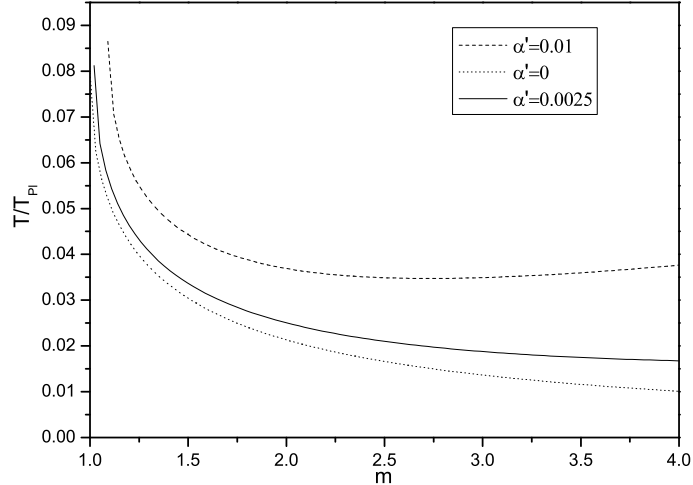}}
  \caption{Temperature of a black hole as a function of mass.
  Temperature and mass are represented in Planck's units. So $m=M/M_{Pl}$
  denotes relation of black hole's mass to the Planck's mass.  Deformation parameters
  are also written in dimensionless form: $\A'=\A L^2_{Pl}$ and $\B'=\h^2\B/L^2_{Pl}$.
  In order to get that the mass of a black hole at the final point of Hawking radiation to be equal
  to the Planck's mass (or $m=1$) when the parameter $\A'=0$ we  set   $\B'=16$.} \label{fig1}
\end{figure}

Minimal uncertainty in position given by relation (\ref{MGUP})
similarly as in case of (\ref{GUP}) prevents us from a temperature
catastrophe. It was proposed that generalized uncertainty
principle (\ref{GUP}) prevents black holes from complete
evaporation in the same way as the standard uncertainty principle
prevents the hydrogen atom from collapse \cite{cavaglia04}. To
make the influence of a minimal uncertainty in momentum clear let
us investigate temperature as a function of $M$. Taking the
derivative $\partial T/\partial M$ and equating it to zero one can
find that temperature has an extremum point when the mass of a
black hole reaches:
\begin{equation}\label{extremal_mass}
M_{ext}=\frac{c^2}{4G\sqrt{\A\left(1-\h^2\A\B\right)}}.
\end{equation}
It is easy to see that at this point temperature of the black hole
takes minimal value:
\begin{equation}\label{minimal_temp}
T_{min}=\frac{c\h\sqrt{\A}}{\pi\sqrt{1-\h^2\A\B}}
\end{equation}
When the black hole's mass is above $M_{ext}$, temperature
(\ref{BH_temp}) increases with the mass rise. When the mass is
below $M_{ext}$ its decreasing leads to increasing of temperature.
Similar mass-temperature dependence was obtained in case of
Schwarzschild-AdS black hole with modified generalized principle
(\ref{MGUP}). We note that minimal temperature is the consequence
of a minimal uncertainty in momentum and it can appear for
different black hole's metrics. Temperature as the function of
mass is shown in Fig.\ref{fig1}.

In the case of higher dimensional space one should use
Schwarzschild-Tangherlini metrics \cite{Tangherlini,Myers}. The
horizon radius of a black hole takes form:
\begin{equation}\label{GR_radius_higher_dimensions}
R_S=\left(\frac{16\pi G_dM}{(d-2)\Omega_{d-2}c^2}\right)^{1/(d-3)}
\end{equation}
Where $d$ is the dimension of space-time, $G_d$ is the
$d$-dimensional gravitational constant and $\Omega_{d-2}$ is the
 surface area of $d-2$-dimensional unit hypersphere.
So the temperature of a $d$-dimensional black hole takes form:
\begin{equation}\label{temp_stringy_high_dim}
T=\frac{(d-3)c}{\pi\h\B}\left[2\left(\frac{16\pi
G_dM}{(d-2)\Omega_{d-2}c^2}\right)^{1/(d-3)}-\sqrt{4\left(\frac{16\pi
G_dM}{(d-2)\Omega_{d-2}c^2}\right)^{2/(d-3)}(1-\h^2\A\B)-\h^2\B}\right]
\end{equation}
As it was pointed out in \cite{scardigli_2} generalized
uncertainty relation (\ref{GUP}) for the black hole in the case of
higher dimensions should be replaced by another one due to the
fact that distances can not be probed below the Schwarzschild
radius (\ref{GR_radius_higher_dimensions}). So according to
\cite{scardigli_2} instead of generalized uncertainty principle
(\ref{GUP}) in $d=4+n$-dimensional space-time one should write:
\begin{equation}\label{MBH_GUP}
\D p\D x\geqslant\frac{\h}{2}(1+\gamma\D p^{\frac{n+2}{n+1}}),
\end{equation}
where $n$ is the number of additional spacelike dimensions. It was
noted \cite{scardigli_2} that at high energies stringy GUP
(\ref{GUP}) can give an uncertainty $\D x$ larger than the size of
a black hole itself. But scenarios with extra dimensions suggest
that all gauge interactions apart from gravity live on $3+1$
dimensional brane \cite{arkani-hamed, arkani-hamed_2, antoniadis,
Randall99}. So modification of GUP (\ref{MBH_GUP}) is valid only
for gravitons, whereas for photons and other SM particles emitted
by a black hole it is necessary to use uncertainty principle
(\ref{GUP}) or (and) (\ref{MGUP}). Taking into account all the
above remarks we can interpret the relation
(\ref{temp_stringy_high_dim}) as some upper bound for temperature
of emitted gravitons.

\section{Entropy and heat capacity of a black hole}
Having calculated black hole's temperature we can find black
hole's entropy using well-known thermodynamical relation:
\begin{equation}
dS=\frac{c^2}{T}dM
\end{equation}
After integration we obtain:
\begin{eqnarray}\label{enthropy_general}
S=\frac{\pi c^3}{4\h
G\A}\left(\ln\Big{|}\frac{4GM-\sqrt{16G^2M^2(1-\h^2\A\B)-\h^2c^4\B}}{\h
c^2\sqrt{\B}}\Big{|}+\right.\nonumber
\\
\left.\sqrt{1-\h^2\A\B}\ln\Big{|}\frac{\sqrt{16G^2M^2(1-\h^2\A\B)-\h^2
c^4\B}+4GM\sqrt{1-\h^2\A\B}}{\h c^2\sqrt{\B}}\Big{|}\right)
\end{eqnarray}
 The last formula reproduces well-known relation for the
Bekenstein-Hawking entropy in the limit $\A,\B\rightarrow 0$. We
can suppose that parameter $\A$ is small in comparing with $\B$ so
we can expand the right hand side of equation
(\ref{enthropy_general}) into the series over a small parameter
$\A$.
\begin{eqnarray}\label{entropy_small_a}
S=\frac{\pi}{4\h c G}\left(8G^2M^2+2GM\sqrt{16G^2M^2-\h^2
c^4\B}-\frac{\h^2c^4\B}{2}\ln\left(\frac{4GM+\sqrt{16G^2M^2-\h^2
c^4\B}}{\h c^2\sqrt{\B}}\right)\right.\nonumber
\\
-\frac{\A}{c^4}\left[64G^4M^4+GM\sqrt{16G^2M^2-\h^2
c^4\B}\left(\frac{\h^2c^4\B}{2}+16G^2M^2\right)\right.\nonumber
\\
\left.\left.+\frac{\h^4c^4\B^2}{2}\ln\left(\frac{4GM+\sqrt{16G^2M^2-\h^2
c^4\B}}{\h c^2\sqrt{\B}}\right)\right]\right)
\end{eqnarray}
In the limit $\A\rightarrow 0$ we reproduce the expression for the
 entropy in the presence of a minimal length \cite {adler}. Expression
(\ref{entropy_small_a}) shows that including $\A$-dependent terms
lead to decreasing of  entropy similarly as we have in the case
with $\B$-terms only.

To calculate the heat capacity of a black hole we use a well-known
thermodynamical relation:
\begin{equation}
C=T\frac{\partial S}{\partial T}=\frac{\partial E}{\partial T}.
\end{equation}
In our case temperature and entropy are represented as functions
of black hole's mass $M$ (or diameter $R$). So we write the heat
capacity as a function of mass $M$.
\begin{equation}\label{heat_capacity_general}
C=\frac{\pi\h
c^3\B}{4G}\frac{\sqrt{16G^2M^2(1-\h^2\A\B)-\h^2c^4\B}}{\sqrt{16G^2M^2(1-\h^2\A\B)-\h^2c^4\B}-4GM(1-\h^2\A\B)}
\end{equation}

\begin{figure}
  \centerline{\includegraphics[scale=1,clip]{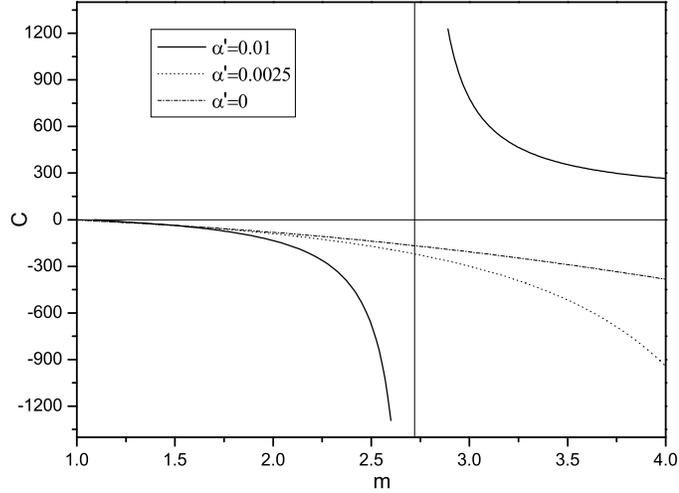}}
  \caption{Heat capacity as a function of mass.}\label{fig2}
\end{figure}

\begin{figure}
  \centerline{\includegraphics[scale=1,clip]{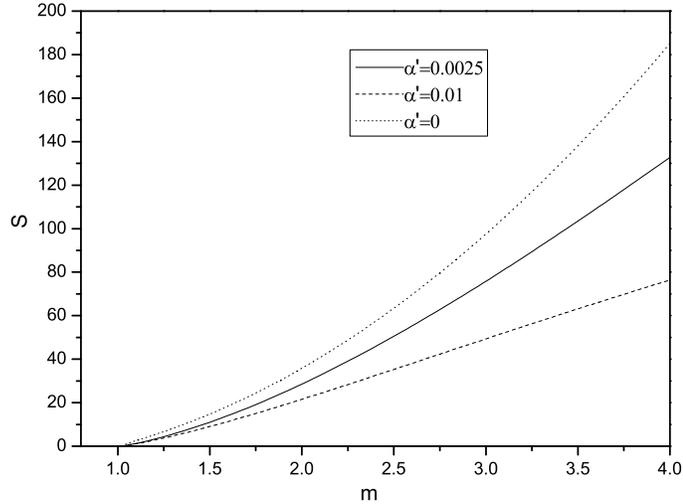}}
  \caption{Entropy as a function of mass}\label{fig3}
\end{figure}

Let us investigate the latter relation in details. It is easy to
make yourself sure that when mass is below $M_{ext}$ heat capacity
is negative and it is equal to zero when the mass reaches
$M_{min}$. When the mass of black hole is above $M_{ext}$ heat
capacity is positive and tends to a finite value when mass goes to
infinity. So $M_{ext}$ is the discontinuity point for the heat
capacity. Heat capacity as a function of mass is represented in
Fig.\ref{fig2}.  Negative heat capacity shows that thermodynamical
system is unstable and tends to decay. Whereas positive heat
capacity makes system stable. So at the extremum point we have
phase transition . One can conclude that the phase transition is
caused by the modified GUP (\ref{MGUP}) but not particular choice
of  black hole's metrics. As it has already been noted modified
GUP (\ref{MGUP}) related to commutation relation (\ref{SDS_alg})
but it is known that the choice of commutation relations that give
the same uncertainty principle is non unique. So modified GUP
(\ref{MGUP}) might be obtained for more general case of
commutation relations. We point out that behaviour of heat
capacity near the phase transition point is similar to behavior of
heat capacity in case of the well-known Hawking-Page phase
transition \cite{Hawking83, Majumdar}. Phase transition also
appears when the modifications of black hole's metrics caused by
nonlocal effects or noncommutative geometry are taken into account
\cite{Nicolini12}. But in that case stable and unstable phases are
exchanged their places in comparison with our result. We also note
that extremum point $M_{ext}$ is an inflection point for entropy.
Entropy as a function of mass is shown in Fig.\ref{fig3}.

\section{Emission rate equation and evaporation time of a black hole}
In this section evaporation of a micro black hole is studied.
Similar problem with GUP was firstly considered in work
\cite{adler}. This topic was investigated in details in
\cite{cavaglia04, scardigli_2}. To examine quantum black hole's
evaporation different approaches are used. Probably the simplest
one is to make use of Stefan-Boltzmann law for the black-body
radiation which is based on an assumption that we have a
thermodynamical system with fixed temperature so we consider
canonical ensemble. We note that the Stefan-Boltzmann law gets new
correction caused by deformation of commutation relations
\cite{cavaglia04, scardigli_2, Chang}. As it was pointed out in
\cite{casdaio, Casadio_2} canonical description for a micro black
hole is adequate  when the energy of emitted particles (e.g.
photons, gravitons) is small in comparison with the energy of a
black hole itself. If the energy of emitted particles is
comparable with the energy of a black hole itself one should use
microcanonical description. When the energy of black hole is large
in comparison with the energy of emitted particles both of them
give the same result. Here we use canonical description,
microcanonical one will be considered elsewhere. We also point out
that statistical mechanics in the case of the deformed commutation
relations was considered in \cite{Fityo}.

Let us examine equilibrium radiation caused by a black hole. We
suppose that that black hole is placed inside a sphere of given
radius $R$ and system is considered under a fixed temperature $T$.
For generality we consider $D$-dimensional case ($D=3+n$). So the
energy of evaporated particles can be calculated as follows
(photons, gravitons):
\begin{equation}\label{energy_blackbody}
E=\frac{g}{(2\pi\h)^D}\int\frac{d^Dxd^Dp}{1+\A x^2+\B
p^2+2\sqrt{\A\B}({\bf x},{\bf p})}\displaystyle
\frac{\varepsilon(p)\Gamma (p)}{e^{\varepsilon(p)/T}-1}
\end{equation}
where $\varepsilon(p)$ is the dispersion relation for emitted
particles, $\Gamma (p)$ is the greybody factor and $g$ is the
degeneracy factor. In the case of photons we put
$\varepsilon(p)=cp$ and $g=2$. The same dispersion relation can be
considered for gravitons but the degeneracy factor is different.
We also suppose that greybody factor is equal to unity
$\Gamma(p)=1$ (blackbody radiation). Let us remind that emission
of gravitons takes place in $d=4+n$ - dimensional bulk spacetime
whereas photons and other gauge particles are emitted on
3+1-dimensional brane.

 We note that suppositions we have made do not
allow to calculate integral exactly (\ref{energy_blackbody}) even
if one of the deformation parameters is absent. To calculate it we
assume that parameters $\A$ and $\B$ are small and develop weight
function under integral into the series over deformation
parameters. So the latter relation can be rewritten in the form:
\begin{equation}
E=\frac{g}{(2\pi\h)^D}\int d^Dxd^Dp (1-\A x^2-\B
p^2-2\sqrt{\A\B}({\bf x},{\bf p}))\displaystyle
\frac{cp}{e^{cp/T}-1}
\end{equation}
After integration we obtain:
\begin{eqnarray}
E=\frac{2g}{(2\sqrt{\pi}c\h)^D}\frac{D!\zeta(D+1)}{\Gamma(D/2)}\left[\left(V-\frac{D^{1+2/D}}{D+2}\left(\frac{\Gamma
\left(D/2\right)}{2\pi^{D/2}}\right)^{2/D}\A
V^{1+2/D}\right)T^{D+1}\right.\nonumber
\\
\left.-\frac{(D+2)!\zeta(D+3)}{D!\zeta(D+1)}\B VT^{D+3}\right]
\end{eqnarray}
where $V=\displaystyle 2\pi^{D/2}R^D/(D\Gamma(D/2))$ is the volume
of $D$-dimensional sphere. It should be noted that the energy of
radiation nonlinearly depends on the volume. In three dimensional
case we find:
\begin{equation}
E=\frac{g}{(2\pi\h)^3}\left[\left(V-\frac{3}{5}\left(\frac{3}{4\pi}\right)^{2/3}\A
V^{5/3}\right)\frac{4\pi^5}{15 c^3}T^4-\frac{32\pi^7}{63c^3}\B
VT^6\right]
\end{equation}
To obtain emission rate equation we suppose that particles are
emitted by sphere with radius $R_S$. So the total energy $dE$
emitted during the period of time $dt$ can be written in the form:
\begin{eqnarray}
\frac{dE}{dt}=-\frac{4gc}{(2\h
c)^D}\frac{D!\zeta(D+1)}{\left(\Gamma(D/2)\right)^2}
\left[\left(R^{D-1}_S-\A
R^{D+1}_S\right)T^{D+1}-\frac{(D+2)!\zeta(D+3)}{D!\zeta(D+1)}\B
R^{D-1}_S\frac{T^{D+3}}{c^2}\right]
\end{eqnarray}
Where $R_S$ is the Schwarzschild radius. For photons in three
dimensional case we obtain:
\begin{equation}\label{emiss_rate_E}
\frac{dE}{dt}=-\frac{4\pi^3}{15c^2\h^3}\left((R^2_S-\A
R^4_S)T^4-\frac{40}{21}\frac{\pi^2\B}{c^2}R^2_ST^6\right)
\end{equation}
In the limit when parameters of deformation tens to zero the last
equation gives ordinary Stefan-Boltzmann law for surface of a
black hole.
 At the final point of Hawking radiation, when the temperature
and Schwarzschild radius reach $T_{final}$ and $R_{min}$
respectively, the emission rate is finite:
\begin{equation}
\frac{dE}{dt}\Big|_{final}=\frac{c^2}{60\h\B(1-\h^2\A\B)^4}\left(\frac{76}{21}+5\h^2\A\B\right).
\end{equation}
Similarly as in the case of GUP \cite{cavaglia04} at the final
point of Hawking radiation emission rate is finite. It was
supposed that when the final stage has been reached, the black
hole evaporates completely by emitting a hard Planck-size quantum
with maximum temperature in a finite period of time proportional
to the Planck's time \cite{cavaglia04}. At the same time heat
capacity (\ref{heat_capacity_general}) tends to zero when the mass
reaches $M_{min}$. It means that black hole cannot exchange heat
with surroundings at the final point of Hawking radiation.

Now using relation (\ref{BH_temp}) and representing Schwarzschild
radius as a function of a black hole's mass we can write equation
for the emission rate of a black hole. Since we have calculated
the energy of emitted radiation up to the first order over
deformation parameters we estimate emission rate up to first order
too. So equation (\ref{emiss_rate_E}) can be written in the form:
\begin{equation}
\frac{dM}{dt}=-\frac{\h c^4}{3840\pi
G^2M^2}\left(1+60\A\frac{G^2M^2}{c^4}+\frac{11}{336}\B\frac{\h^2
c^4}{G^2M^2}\right)
\end{equation}

\begin{figure}
  \centerline{\includegraphics[scale=1,clip]{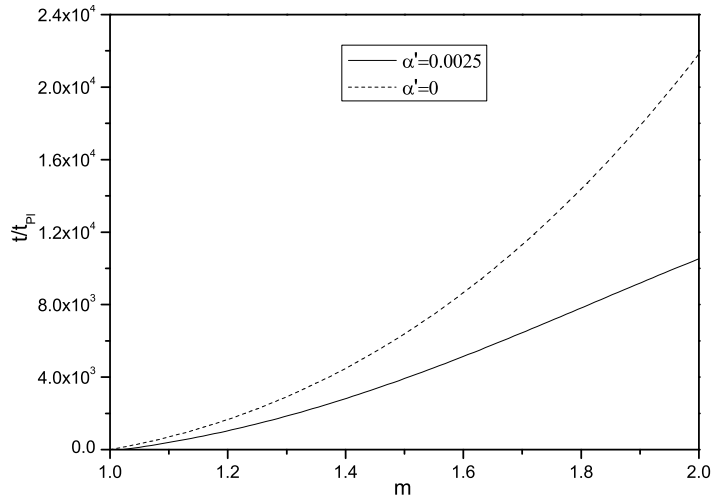}}
  \caption{Evaporation time as a function of mass}\label{fig4}
\end{figure}

Having integrated the last equation we obtain simple expression
for evaporation time of a black hole:
\begin{equation}\label{time_of_evapor}
\frac{1}{3}(M^3_{min}-M^3)-12\frac{\A
G^2}{c^4}(M^5_{min}-M^5)-\frac{11}{336}\frac{\h^2\B
c^4}{G^2}(M_{min}-M)=-\frac{\h c^4}{3840\pi G^2}t
\end{equation}
One can see that modification of generalized uncertainty principle
(\ref{MGUP}) leads to further decrease in black hole's evaporation
time in comparison with GUP (\ref{GUP}). Evaporation time as a
function of mass is shown in Fig.\ref{fig4}

\section{Conclusions}
We have considered microscopic black hole with modified GUP
(\ref{MGUP}) leading to the appearance of a minimal length as well
as minimal momentum. Uncertainty relation allows one to obtain
thermodynamical functions of a black hole in a very simple way.
Therefore it was the main motivation to investigate the black
hole's thermodynamics under a bit more general assumption such as
generalized uncertainty principle \cite{adler, cavaglia04}.
Modification of GUP (\ref{MGUP}) gives rise to some new features
in black hole's thermodynamics. Similarly as in case of GUP
(\ref{GUP}) black hole has finite final temperature which is
caused by a minimal uncertainty in position. However, in contrast
to GUP (\ref{GUP}) modified GUP (\ref{MGUP}) leads also to minimal
temperature (\ref{minimal_temp}). Similar result was firstly
obtained in \cite{Park08} but there Schwarzschild-AdS black hole's
metrics was used and a specific choice for one of the deformation
parameters was made. This minimal temperature causes important
influence on black hole's thermodynamics. The point of minimal
temperature is a point of discontinuity for heat capacity. As we
have already pointed out when the mass of a black hole is below
$M_{ext}$ (\ref{extremal_mass}) heat capacity is negative and it
tells us about thermodynamical instability. So in this case black
hole tends to decay. When the black hole mass is above $M_{ext}$
heat capacity is positive and black hole is thermodynamically
stable. We also note that although modified GUP (\ref{MGUP}) is
related to Snyder-de Sitter commutation relations it can obtained
for a more general kind of commutation relations. We can conclude
that phase transition which appears due to the presence of minimal
momentum and it is possible for different black holes metrics.

We also investigated thermal radiation of a Schwarzschild black
hole. We obtained the emission rate equation which is based on
modified Stefan-Boltzmann law. Then we used it to calculate the
evaporation time of a Schwarzschild black hole. In comparison with
ordinary GUP (\ref{GUP}) modified GUP (\ref{MGUP}) makes
evaporation time of a black hole shorter. We also point out that
calculations of evaporation time are valid only for unstable phase
of a black hole when its mass is below $M_{ext}$.

\section{Appendix}
In order to obtain correct relation for the black body radiation
spectrum we should modify Liouville theorem of the classical
mechanics caused by a deformation of commutation relations. So we
have to find an element of phase-space volume that is invariant
under time evolution. To do this let us write classical equations
of motion supposing that position and momentum coordinates are
obeyed the classical variant of the algebra (\ref{SDS_alg}). In
order to obtain the classical commutation relations the standard
procedure is used:
\begin{equation}
\frac{1}{i\h}[\hat{A},\hat{B}]\Rightarrow \{A,B\}
\end{equation}
Under this assumption deformed commutation relation
(\ref{SDS_alg}) takes following form:
\begin{eqnarray}\label{SDS_brackets}
\{x_i,p_j\}=\delta_{ij}+\A x_ix_j+\B p_ip_j+2\sqrt{\A\B}x_jp_i.
\end{eqnarray}
As it has already been pointed out in the classical case
commutation relation (\ref{SDS_brackets}) forms a well defined
algebra. Other two commutation relations takes form
\cite{Banerjee, mignemi}:
\begin{eqnarray}
{\{x_i,x_j\}=\B J_{ij}};\quad \{p_i,p_j\}=\A J_{ij},\nonumber
\end{eqnarray}
where $J_{ij}$ are the components of angular momentum.

Hamilton's equations for the time derivatives of position and
momentum read:
\begin{eqnarray}
\dot{x}_i=\{x_i,H\}=\{x_i,x_j\}\frac{\partial H}{\partial
x_j}+\{x_i,p_j\}\frac{\partial H}{\partial p_j},
\\
\dot{p}_i=\{p_i,H\}=\{p_i,x_j\}\frac{\partial H}{\partial
x_j}+\{p_i,p_j\}\frac{\partial H}{\partial p_j}.\nonumber
\end{eqnarray}
Considering evolution of a system during an infinitesimal period
of time we obtain:
\begin{eqnarray}
x'_i=x_i+\delta x_i;
\\
p'_i=p_i+\delta p_i.
\end{eqnarray}
Here
\begin{eqnarray}
\delta x_i=\dot{x}_i\delta t=\left(\{x_i,x_j\}\frac{\partial
H}{\partial x_j}+\{x_i,p_j\}\frac{\partial H}{\partial
p_j}\right)\delta t;
\\
\delta p_i=\dot{p}_i\delta t=\left(\{p_i,x_j\}\frac{\partial
H}{\partial x_j}+\{p_i,p_j\}\frac{\partial H}{\partial
p_j}\right)\delta t.
\end{eqnarray}
After infinitesimal evolution an element of phase-space volume
changes:
\begin{equation}\label{phase_sp_volume}
d^Dx'd^Dp'=\Big|\frac{\partial(x'_1,\ldots,x'_D,p'_1,\ldots,
p'_D)}{\partial(x_1,\ldots,x_D,p_1,\ldots, p_D)}\Big|d^Dxd^Dp
\end{equation}
For generality we consider here $D$-dimensional case. For the
derivatives we have:
\begin{eqnarray}
\frac{\partial x'_i}{\partial x_j}=\delta_{ij}+\frac{\partial
\dot{x}_i}{\partial x_j}\delta t;\quad \frac{\partial
x'_i}{\partial p_j}=\frac{\partial \dot{x}_i}{\partial p_j}\delta
t;
\\
\frac{\partial p'_i}{\partial x_j}=\frac{\partial
\dot{p}_i}{\partial x_j}\delta t; \quad \frac{\partial
p'_i}{\partial p_j}=\delta_{ij}+\frac{\partial \dot{p}_i}{\partial
p_j}\delta t.
\end{eqnarray}
We calculate the Jacobian in the relation (\ref{phase_sp_volume})
up to the first order over the infinitesimal time translation
$\delta t$. So under this approximation the Jacobian can be
written in the form:
\begin{equation}
J=\Big|\frac{\partial(x'_1,\ldots,x'_D,p'_1,\ldots,
p'_D)}{\partial(x_1,\ldots,x_D,p_1,\ldots,
p_D)}\Big|=1+\left(\frac{\partial\dot{x}_i}{\partial
x_i}+\frac{\partial\dot{p}_i}{\partial p_i}\right)\delta t
\end{equation}
 So we calculate:
\begin{eqnarray}
\frac{\partial\dot{x}_i}{\partial
x_i}+\frac{\partial\dot{p}_i}{\partial
p_i}=\frac{\partial}{\partial x_i}\left(\{x_i,x_j\}\frac{\partial
H}{\partial x_j}+\{x_i,p_j\}\frac{\partial H}{\partial
p_j}\right)+\frac{\partial}{\partial
p_i}\left(\{p_i,x_j\}\frac{\partial H}{\partial
x_j}+\{p_i,p_j\}\frac{\partial H}{\partial p_j}\right)\nonumber
\\
=\frac{\partial}{\partial x_i}\big[\{x_i,x_j\}\big]\frac{\partial
H}{\partial x_j}+\{x_i,x_j\}\frac{\partial^2 H}{\partial
x_i\partial x_j}+\frac{\partial}{\partial
x_i}\big[\{x_i,p_j\}\big]\frac{\partial H}{\partial
p_j}+\{x_i,p_j\}\frac{\partial^2 H}{\partial x_i\partial
p_j}+\nonumber
\\
\frac{\partial}{\partial p_i}\big[\{p_i,x_j\}\big]\frac{\partial
H}{\partial x_j}+\{p_i,x_j\}\frac{\partial^2 H}{\partial
p_i\partial x_j}+\frac{\partial}{\partial
p_i}\big[\{p_i,p_j\}\big]\frac{\partial H}{\partial
p_j}+\{p_i,p_j\}\frac{\partial^2 H}{\partial p_i\partial
p_j}\nonumber
\\
=2\left(\A x_k\frac{\partial H}{\partial p_k}-\B p_k\frac{\partial
H}{\partial x_k}+\sqrt{\A\B}\left[p_k\frac{\partial H}{\partial
p_k}-x_k\frac{\partial H}{\partial x_k}\right]\right)
\end{eqnarray}
For the element of phase space volume we have:
\begin{equation}\label{phase_sp_volume}
d^Dx'd^Dp'=d^Dxd^Dp\left[1+2\left(\A x_k\frac{\partial H}{\partial
p_k}-\B p_k\frac{\partial H}{\partial
x_k}+\sqrt{\A\B}\left[p_k\frac{\partial H}{\partial
p_k}-x_k\frac{\partial H}{\partial x_k}\right]\right)\delta
t\right]
\end{equation}
Let us consider:
\begin{eqnarray}\label{weight_multipl}
1+\A x'^2+\B p'^2+2\sqrt{\A\B}({\bf x}',{\bf p}')=1+\A(x_i+\delta
x_i)^2+\B(p_i+\delta p_i)^2\nonumber
\\
+2\sqrt{\A\B}(x_i+\delta x_i,p_i+\delta p_i)\simeq 1+\A x^2+\B
p^2+2\sqrt{\A\B}({\bf x},{\bf p})\nonumber
\\
+2(\A(x_i,\dot{x}_i)+\B(p_i,\dot{p}_i)+\sqrt{\A\B}[(x_i,\dot{p}_i)+(p_i,\dot{x}_i)])\delta
t=(1+\A x^2+\B p^2\nonumber
\\
+2\sqrt{\A\B}({\bf x},{\bf p}))\left[1+2\left(\A x_k\frac{\partial
H}{\partial p_k}-\B p_k\frac{\partial H}{\partial
x_k}+\sqrt{\A\B}\left[p_k\frac{\partial H}{\partial
p_k}-x_k\frac{\partial H}{\partial x_k}\right]\right)\delta
t\right]
\end{eqnarray}
Making use of relations (\ref{phase_sp_volume}) and
(\ref{weight_multipl}) we conclude that the following weighted
phase space volume is invariant under infinitesimal time
translations:
\begin{equation}\label{weighted_volume}
\frac{d^Dxd^Dp}{1+\A x^2+\B p^2+2\sqrt{\A\B}({\bf x},{\bf p})}
\end{equation}

\end{document}